\documentclass[aps,prb,preprint]{revtex4-1}
\usepackage{graphicx, bm, epsfig}
\begin{document}
\title{Discrete Charge Dielectric Model of Electrostatic Energy}
\author{Tim LaFave Jr.}

\affiliation{University of Texas at Dallas, Department of Electrical Engineering, Richardson, TX 75080}

\begin{abstract}
Studies on nanoscale materials merit careful development of an electrostatics model concerning discrete point charges within dielectrics. The discrete charge dielectric model treats three unique interaction types  derived from an external source: Coulomb repulsion among point charges, direct polarization between point charges and their associated surface charge elements, and indirect polarization between point charges and surface charge elements formed by other point charges. The model yields the potential energy, $U(N)$, stored in a general $N$ point charge system differing from conventional integral formulations, $1/2\int{\bm E}\cdot{\bm D}dV$ and $1/2\int\rho\Phi dV$, in a manner significant to the treatment of few electron systems.
\end{abstract}
\maketitle

\section{Introduction}
The most common conceptual approach to evaluate the total electrostatic potential energy stored in a dielectric containing free point charges is the method of charge assembly. Textbook treatments\cite{griffiths1999, jackson1999, vanderlinde1993, purcell1985, schwartz1987, reitz1993, sadiku1995, becker1982, visscher1988} replace discrete point charges within a dielectric with a continuous free charge density, $\rho$. Energy stored in a dielectric containing $\rho$ is easily evaluated by integration over the volume of the dielectric in which the smallest increment of charge, $dq$, is zero. Though this is a reasonable approximation when $\rho$ consists of a large number of point charges,\cite{vanderlinde1993} it fails to reflect the physical nature of point charge configurations commonly found in nanoscale systems today. Moreover, the imposition of a {\it continuous} charge density prohibits the practical formation of any realistic charge distributions composed of infinitesimal, non-existent charge increments. The present model is set apart from these approaches in that a summation of interaction terms is obtained rather than integral approximations based on the interaction of $\rho$ with a macroscopic total potential, $\Phi$,\cite{griffiths1999}

\begin{eqnarray}
U(\rho) = \frac{1}{2} \int_V \rho \Phi dV\label{eq:lafave-rhophi}
\end{eqnarray}

{\noindent}or an electric field formulation,\cite{jackson1999}

\begin{eqnarray}
U({\bm E}) = \frac{1}{2} \int_V {\bm E}\cdot {\bm D} dV.\label{eq:lafave-EdotD}
\end{eqnarray}

The objective of the present model is to account for the interaction of each new point charge to the system with the sum of all electrostatic potentials due to every pre-existing charge and charges formed during the assembly process without assumption of an infinitesimal charge increment, $dq$, to reflect the natural existence of a fundamental charge increment, $e$. 

The discrete charge dielectric model has been used for numerical evaluations in recent studies concerning the capacitance of few-electron spherical quantum dots\cite{LaFaveJr2009-791, LaFaveJr2008-1269, LaFaveJr2008-617, LaFaveJr-Dissertation} and ``meta-atoms''\cite{fiddy2010}, cited as leading to charge carrier localization near quantum dot surfaces\cite{PhysRevB.76.045401} and played a central role in the discovery of a classical electrostatic fingerprint of atomic structure.\cite{LaFaveJr2008-617, LaFaveJr-Dissertation} The present model represents a redevelopment and generalization of a previously published interactions picture initially used to evaluate the electrostatic energy of a silicon sphere containing one or two electrons embedded in a uniform silicon dioxide environment.\cite{Tsu2005, PhysRevB.45.14150} Exclusion of one-half of the ``self-polarization'' interaction energy term in the original picture is shown to be inconsistent with the current formulation in which a more appropriate description, ``direct polarization,'' is used to both eliminate confusion with self-interaction or self energy -- interaction of a point charge with its own potential -- and to differentiate it from indirect polarization interactions (``polarizations'' in the original picture) which are not directly involved in the polarization of the dielectric.

A practical definition of stored energy is that delivered to the system by an external agent acting on point charges during assembly and must be equal to that recovered by an external conservative energy source when the system is dismantled. This is necessary such that the dielectric is returned to its original, neutral state. The amount of energy displaced by polarization of atomic or molecular constituents of the dielectric must likewise be restored to its {\it internal} conservative energy source (atoms or molecules) but does not contribute to the energy stored in the system as it already exists within it.

In the following, the electrostatic interaction between two point charges of like sign is carefully described. Seemingly trivial, the development is exceedingly instructive toward a physical intuition and constructs a framework in which the other types of interaction are developed. This approach identifies the mutual distribution of interaction energy between charges and categorizes interactions as directly resulting from work done by an external agent or a response of charges already in the system to new charge arrivals. 

The electrostatic constraint of a single point charge to a dielectric object is examined. Formation of a bound surface charge element at the dielectric interface is the result of direct polarization, an example of the charged system's response to a newly introduced point charge. In the process of introducing a second point charge to the system, the point charges interact directly with each other, and a new bound surface charge is formed by direct polarization. Additionally, the external agent must do work on the new point charge against the electrostatic potential due to the bound surface charge element formed exclusively by the first point charge. This represents an indirect polarization in which energy is mutually shared between the two point charges as the first point charge is likewise subjected to the potential due to the additional bound surface charge element formed by the second point charge. The latter interaction is a necessary response to the formation of an additional bound charge. The energy associated with this response can only derive from work done by the external agent. For completeness, interactions among bound charges are arguably intrinsic to the system and have no direct source of energy as may only be supplied by the external agent. These interactions are therefore explicitly taken into account within the framework of the discrete charge dielectric model during assembly.

For few-electron systems commonly found in nanoscale studies the generalized summation expression differs from Eqs.~\ref{eq:lafave-rhophi} and \ref{eq:lafave-EdotD} by up to a factor of $1/2$ ($N=1$). This result is not unexpected as it is consistent with a similar charge-by-charge derivation\cite{PhysRevB.52.10737} of the so-called ``quantum'' capacitance expression, $C_{\textrm{\small{quant}}} = Q^2/U$, which differs from the familiar ``classical'' capacitance expression, $C_{\textrm{\small{class}}} = Q^2/2U$, by a factor of 1/2. Similarly, a $1/2$ capacitance disparity between ``open'' (isolated) and ``closed'' (connected) single electron devices is known.\cite{Tsu2005} The sources of these disparities are discussed and represent one-half the total direct polarization energy in the present model.

In the other extreme, as $N$ grows large, the direct polarization energy term becomes a negligible fraction of the total energy as required for the approximation of $dq\to 0$. Hence, the discrete charge dielectric model is applicable to nanoscale systems in which the behavior of few free electrons dominates the electronic properties of a material, as well as many-electron systems limited by computing capacity and other practical considerations. The generalized expression may be numerically minimized to obtain a final electrostatic configuration for any $N$-electron system in satisfaction of necessary boundary conditions and the shape and size of the dielectric.

\section{Discrete Charge Dielectric Model}

The electrostatic force acting on a point charge, $q$, by an electric field, $\mathbf{E}$, due to all other sources of charge in a system, is given by $\mathbf{F} = q\mathbf{E}$. The work done to displace the point charge by a small distance, $dx$, is $dW = -\mathbf{F}\cdot d\mathbf{x}$ such that the total work done to move $q$ from a non-interacting location, $r_n$, to an interacting location, $r_i$, in the vicinity of a charged system is given by

\begin{eqnarray}\label{eq:lafave-work}
W &=& \int_{r_n}^{r_i} dW = - q \int_{r_n}^{r_i} \mathbf{E}\cdot d\mathbf{x}\nonumber\\
&=& q \left[ \Phi(r_i) - \Phi(r_n)\right]\nonumber\\
&=& q\Phi(r_i).\label{eq:lafave-interaction}
\end{eqnarray}

{\noindent}Here, $\Phi$ is the electrostatic potential due to all sources of charge present in the system before and during the displacement of $q$.

A single point charge in the absence of other charges is considered to have no stored energy as no work is done to move the point charge around. A point charge, $q_i$, interacting with its own electrostatic potential, 

\begin{eqnarray*}
\phi(q_i, r_i) = \frac{q_i}{4\pi \varepsilon_0}\frac{1}{|{\bm r}_i - {\bm r}_i|}
\end{eqnarray*}

{\noindent}at its location, $r_i$, is infinite in magnitude. This self-interaction energy is excluded from evaluation of the total stored energy as the external agent is not involved.

\subsection{Coulomb Interaction}
To evaluate the ``Coulomb interaction''\cite{PhysRevB.45.14150} energy between $q_1$ and $q_2$ in free space having the same sign consider an external agent acting on $q_2$ in the direction of $q_1$. As $q_2$ approaches $q_1$ retreats to infinity as there are no physical constraints acting on it. If there are physical constraints, an evaluation of the energy associated with the constraining forces must be made. This concept is central to the discrete charge dielectric model, and dismisses the unnecessary prescription of an initially ``fixed'' charge configuration conventionally implied in the electrostatics literature. Hence, the {\it response} of the pre-existing charge configuration to new charge arrivals is paramount. This perspective readily facilitates the introduction of new charges to the system.

If $q_1$ is found within a dielectric characterized by dielectric constant, $\varepsilon$, the displacement of $q_1$ is the result of $q_1$ interacting with the potential due to $q_2$. This interaction represents the response of $q_1$ to the arrival of $q_2$, but may only derive its energy indirectly from the work done on $q_2$ by the external agent. As the interaction terms are equal by symmetry the energy supplied by the external agent acting on $q_2$ against $\phi(q_1,r_2)$, 

\begin{eqnarray}
U_{\textrm{\small{C}}}(q_1,q_2) &=& q_2 \phi(q_1, r_2)\nonumber\\
&=& q_2 \frac{q_1}{4\pi\varepsilon_0\varepsilon}\frac{1}{|{\bm r}_1 - {\bm r}_2|}.\label{eq:lafave-coulomb}
\end{eqnarray}

{\noindent}is mutually shared between $q_1$ and $q_2$. The notation $\phi(q_1,r_2)$ denotes the source of electrostatic potential and location at which it acts on $q_2$. The notation $U_C(q_1,q_2)$ denotes the charges with which energy is mutually shared.

The Coulomb interaction is subdivisible into two interactions: 1) a response of $q_1$ indirectly supplied energy through 2) direct Coulomb interaction of an external agent acting upon $q_2$ during the assembly process. The latter is the sole source of stored energy involved in the Coulomb interaction. 

Fig.~\ref{fig:lafave-1}a schematically depicts a configuration of two point charges and resulting dipole distributions of net-polarized bound surface charge elements, $\sigma_1$ and $\sigma_2$, in a dielectric. Fig.~\ref{fig:lafave-1}b shows the electrostatic interactions necessary for a complete evaluation of the total stored energy. The Coulomb interaction, $U_C$, is schematically shown with mutually-directed arrows between point charges $q_1$ and $q_2$.

\subsection{Direct Polarization}
Consider the process by which $q_1$ is introduced into the dielectric in the previous discussion. The dielectric is initially electrically neutral as the positive and negative charges within the material are arranged such that there is no net accummulation of charge. To this system is introduced a negative point charge, $q_1$, by an external agent. The external agent must supply energy in its direct interaction with $q_1$ to act indirectly against the restorative nature of the charged atomic or molecular constituents of the dielectric in a process of polarization. Due to the dielectric disparity at the boundary, a net negative region of charge is formed. The result is the formation of a unique bound surface charge element, $\sigma_1$. If $q_1$ were removed from the system, the restorative nature of the atomic or molecular constituents act to return the displaced energy to its internal source. This process is representative of a conservative internal energy source. 

The direct polarization energy,

\begin{eqnarray}\label{eq:lafave-dir}
U_{\textrm{\small{dir}}}(q_1,\sigma_1) = q_1 \phi(\sigma_1, r_1),
\end{eqnarray}

{\noindent}is mutually shared between $q_1$ and $\sigma_1$, contributing to the energy stored in the free point charge distribution, $U_{\textrm{\small{free}}}$, and the bound surface charge, $U_{\textrm{\small{bound}}}$, respectively. This is schematically shown in Fig.~\ref{fig:lafave-1}b with mutually-directed arrows for $U_{\textrm{\small{dir}}}$. The total energy stored in a single point charge system is given solely by the direct polarization interaction, 

\begin{eqnarray}
U(1) = U_{\textrm{\small{dir}}}(q_1,\sigma_1).
\end{eqnarray}

\subsection{Indirect Polarization}
\begin{figure}[ht!]
\includegraphics[width=10cm]{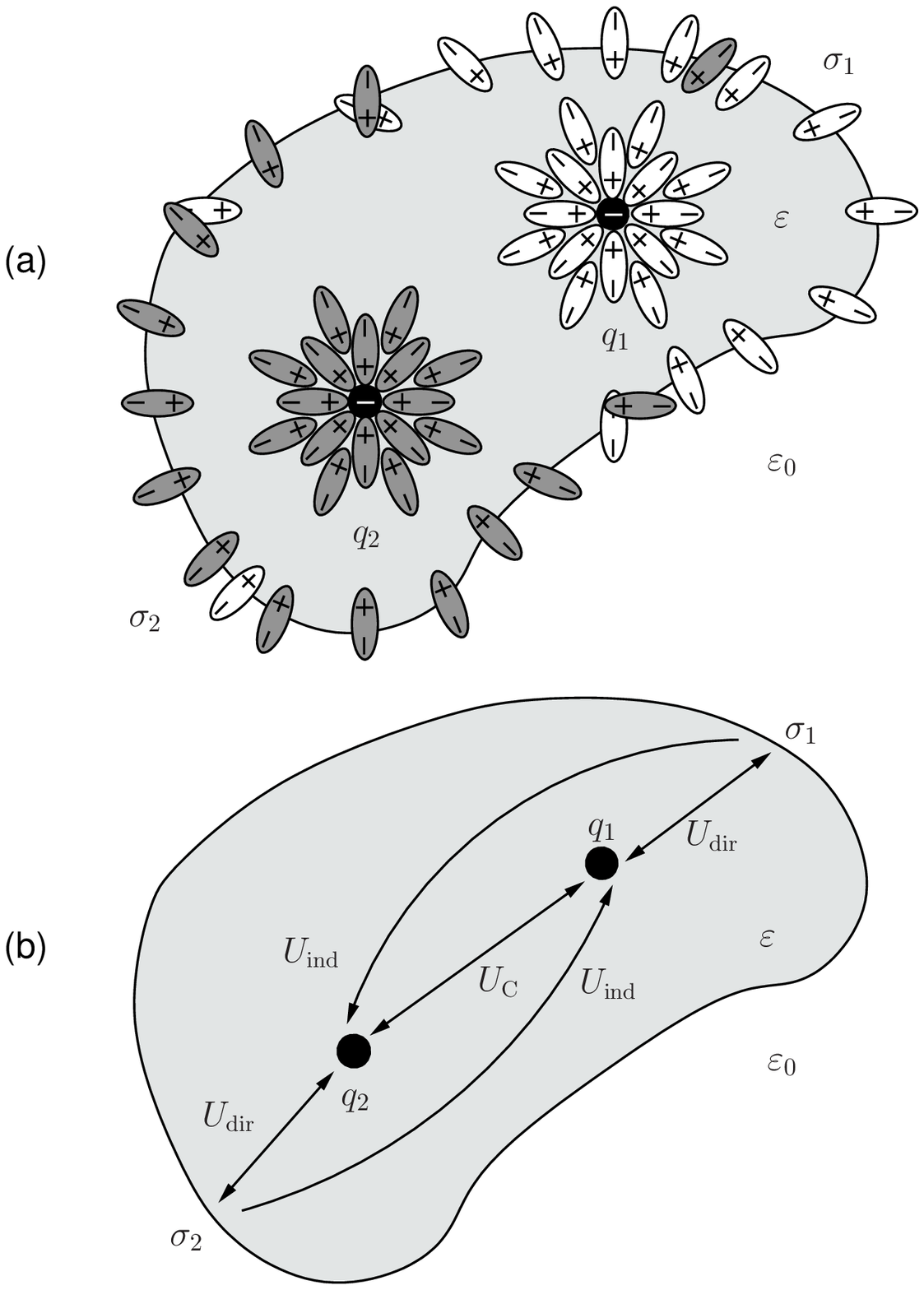}
\caption{Discrete Charge Dielectric Model a) Two point charges, $q_1$ and $q_2$, repel each other and polarize the dielectric such that net bound surface charge elements, $\sigma_1$ and $\sigma_2$, are formed. b) Coulomb repulsion, $U_C$, is schematically represented by a mutually-directed arrow between $q_1$ and $q_2$ indicating mutual energy distribution. Likewise, for direct polarizations, $U_{\textrm{\small{dir}}}$, energy is mutually shared between each point charge, $q_i$, and its respectively polarized surface charge element, $q_i$. As $\sigma_i$ are exlusively dependent upon respective $q_i$ energy derived from indirect polarizations, $U_{\textrm{\small{ind}}}$, are mutually shared between $q_1$ and $q_2$, denoted by singly-directed arrows.}
\label{fig:lafave-1}
\end{figure}

Introduction of a second point charge, $q_2$, to the system (Fig.~\ref{fig:lafave-1}a) involves work done on $q_2$ by an external agent against the electrostatic potential of $q_1$, $\sigma_2$ and $\sigma_1$. Interactions with potential due to $q_1$ and $\sigma_2$ are respectively given by Eq.~(\ref{eq:lafave-coulomb}) and 

\begin{eqnarray}\label{eq:lafave-dir2}
U_{\textrm{\small{dir}}}(q_2,\sigma_2) = q_2 \phi(\sigma_2, r_2).
\end{eqnarray}

{\noindent}Interaction with potential due to $\sigma_1$ results in the {\it indirect} polarization of $\sigma_1$ by $q_2$ given by 

\begin{eqnarray}\label{eq:lafave-ind}
U_{\textrm{\small{ind}}}(q_1,q_2) = q_2\phi(\sigma_1, r_2).
\end{eqnarray}

{\noindent}The bound surface charge element $\sigma_1$ is {\it exclusively} dependent upon the magnitude and location of $q_1$. Therefore, interaction of $q_2$ with $\phi(\sigma_1,r_2)$ does not {\it directly} redistribute $\sigma_1$. Instead, the introduction of $q_2$ results in the formation of $\sigma_2$ which presents a potential that displaces $q_1$. The displacement of $q_1$ results in a redistribution of $\sigma_1$ by direct polarization. Hence, $\sigma_1$ is indirectly polarized by $q_2$. 

Energy associated with the interaction of $q_1$ with $\phi(\sigma_2,r_1)$ must have a source. Work done on $q_2$ is the sole source of energy. The source of energy associated with the indirect polarization of $\sigma_2$ by $q_1$ is found by distributing energy among the entire system of charges. Tracing through Fig.~\ref{fig:lafave-1}b, the interaction energy associated with $q_2$ acting against the potential due to $\sigma_1$ remains associated with $q_2$ (hence, the arrow from $\sigma_1$ to $q_2$). This energy may be distributed along two interaction ``paths'': 1) through Coulomb interaction with $q_1$, and 2) through direct polarization with $\sigma_2$. Energy associated with the indirect polarization of $\sigma_2$ by $q_1$ remains associated with $q_1$, explicitly denoted in Eq.~\ref{eq:lafave-ind}, as $\sigma_2$ is exclusively dependent on $q_2$.

In satisfaction of interaction symmetry, energy is redistributed throughout the charge system (Fig.~\ref{fig:lafave-1}b) such that energy resulting from the indirect polarization of $\sigma_1$ by $q_2$ is mutually shared by $q_1$ and $q_2$. The total indirect polarization energy contributes to $U_{\textrm{\small{free}}}$. Any contributions made to $U_{\textrm{\small{bound}}}$ are taken into full account through respective direct polarizations due to mutual displacements of $q_1$ and $q_2$ by $\sigma_2$ and $\sigma_1$, respectively. 

The energy stored in a dielectric containing two point charges is then 

\begin{eqnarray}\label{eq:lafave-u2}
U(2) = U_C(q_1,q_2) + U_{\textrm{\small{ind}}}(q_1,q_2) + \sum_{i=1}^2 U_{\textrm{\small{dir}}}(q_i,\sigma_i).
\end{eqnarray}

\subsection{Surface Charge Elements}
For completeness, electrostatic interactions among surface charge elements, $\sigma_i$, have no direct source of energy and must instead draw upon energy provided indirectly by the respective direct polarizations as-evaluated. Noted previously, the magnitude and distribution of each $\sigma_i$ is {\it exclusively} dependent upon the magnitude and location of its associated $q_i$. Similarly, energy associated with the indirect polarization interaction $q_1 \phi(\sigma_2,r_1)$ is not evaluated directly but draws mutually upon the indirect polarization interaction $q_2 \phi(\sigma_1, r_2)$. Additionally, careful treatment of Coulomb interaction energy between two point charges yields only one-half of the {\it apparent} sum of energy resulting from both interactions. Explicitly

\begin{eqnarray}
U_C(q_1,q_2) = \frac{1}{2} \left[ q_1\phi(q_2,r_1) + q_2 \phi(q_1, r_2) \right]
\end{eqnarray}

{\noindent}is equivalent to Eq.~\ref{eq:lafave-coulomb}. Eq.~\ref{eq:lafave-coulomb} implicitly takes into account the interaction $q_1\phi(q_2,r_1)$ as $q_2\phi(q_1,r_2)$ is an indirect internal response.

Hence, interactions among $\sigma_i$ are explicitly taken into account by evaluation of all work done by an external agent acting upon charges assembled to the dielectric within the framework of the discrete charge dielectric model. 

\subsection{$N$ Electrons}

As each new electron is assembled to the system its interaction with the electrostatic potential due to all charges are taken into full account by three types of interaction. The model may be generalized to a dielectric containing $N$ electrons. The total stored electrostatic energy is given by generalization of Eq.~\ref{eq:lafave-u2},

\begin{eqnarray}\label{eq:lafave-un}
U(N) = \frac{1}{2} \sum_{i=1}^N \sum_{j\ne i}^{N} U_{\textrm{\small{C}}}(q_i,q_j) + \frac{1}{2} \sum_{i=1}^{N} \sum_{j\ne 1}^{N} U_{\textrm{\small{ind}}}(q_i,q_j) + \sum_{i=1}^N U_{\textrm{\small{dir}}}(q_i,\sigma_i)
\end{eqnarray}

{\noindent}where mutual energy distribution with charge pairs within each interaction is explicitly denoted.

\section{Discussion}
The total energy stored in a dielectric containing discrete point charges must be equal to the amount of energy delivered to the system in the form of work done by an external agent during charge assembly. This energy is evaluated by taking into account the interaction of each discrete point charge with the total electrostatic potential arising from all charges in the system both before and during the assembly process. Full partitioning of energy during assembly is provided by dividing interactions into three types: Coulomb interactions among point charges, direct polarization between each point charge and its exclusively bound surface charge element, and indirect polarization resulting from the interaction of each new point charge with bound surface charge elements formed by other point charges. 

Subdividing each type of interaction into those contributing to the energy stored due to work directly done by an external agent and those that must be indirectly satisfied by response to new charge arrivals, partitioning of energy is completely comprehended. By the symmetric relationship between interacting charges, mutual distribution of energy is observed. The energy may be broadly associated with the free point charge distribution, $U_{\textrm{\small{free}}}$, and the bound surface charge, $U_{\textrm{\small{bound}}}$.\cite{griffiths1999} Energy resulting from Coulomb interactions, indirect polarization interactions and one-half of direct polarization interactions contribute to $U_{\textrm{\small{free}}}$. One half of the energy resulting from direct polarization interactions contribute to $U_{\textrm{\small{bound}}}$. The total bound charge, $\sigma$, is comprised of contributions from bound surface charge elements, $\sigma_i$. Interactions among $\sigma_i$ are explicitly taken into account in the present model, having acknowledged them as internal responses with no direct sources of energy beyond those discussed.

Unique to the discrete charge dielectric model is its treatment of discrete point charges in dielectrics leading to a general summation, Eq.~\ref{eq:lafave-un}, in place of conventional integral formulations, Eqs.~\ref{eq:lafave-rhophi} and \ref{eq:lafave-EdotD}. The difference between the present formulation and integral formulations is the factor of $1/2$ in the latter. The last term of Eq.~\ref{eq:lafave-un} is {\it not} halved as both the energy required to form surface charge elements and the energy associated with point charges due to interaction with resulting surface charge elements must be added to the system. 

The amount of work done in the formation of each $\sigma_i$ is given by\cite{LaFaveJr2008-617}

\begin{eqnarray}
W = \int_0^{q_i} \phi(\sigma_i, r_i) dq = \frac{q_i\phi(\sigma_i, r_i)}{2}
\end{eqnarray}

{\noindent}since $\phi(\sigma_i, r_i)\propto q$, and $\sigma_i$ is justifiably formed by charge increments $dq\to 0$. One-half of each direct polarization interaction $q_i\phi(\sigma_i, r_i)$ is consumed in the process and stored in $\sigma_i$, while the remaining half is associated with each $q_i$ (part of $U_{\textrm{\small{free}}}$). The bound energy, $U_{\textrm{\small{bound}}}$, must not be removed by the invention of a negative sign on a ``spring energy'' term, $U_{\textrm{\small{spring}}}$\cite{griffiths1999} nor merely set to zero except in the restrictive case of fixed external source charges of an electric field.\cite{jackson1999} For this reason there is no factor of $1/2$ on the third term of Eq.~\ref{eq:lafave-un}.

In the few-electron extreme of $N=1$, the discrete charge dielectric model yields twice the total stored energy given by textbook formulations. This result is anticipated as it leads to an expression of capacitance\cite{LaFaveJr2009-791, LaFaveJr2008-1269, LaFaveJr2008-617, LaFaveJr-Dissertation} identical to that proposed in the past couple decades for a so-called ``quantum'' capacitance expression, $C_{\textrm{\small{quant}}} = Q^2/U$ that differs from the conventional expression by a factor of $1/2$. An explanation of this disparity is immediately identified with the presumption of $dq\to 0$ in the derivation of $C_{\textrm{\small{class}}}$ given by evaluation of the work done to ``charge'' a dielectric, 

\begin{eqnarray}
W &= \int dW &= \int_0^Q \phi dq\nonumber\\
&=  \int_0^Q \frac{q}{C} dq &= \frac{Q^2}{2C}.
\end{eqnarray}

{\noindent}Alternatively, derivation using a discrete charge element, $\Delta q = e$ leads to $W = Q^2/C$ with $Q = Ne$ using a charge-by-charge derivation similar to that leading to a ``quantum'' capacitance\cite{PhysRevB.52.10737}.

Additionally, the one-half disparity between the present model and conventional formalism is also observed between isolated and connected single electron devices.\cite{Tsu2005} In this context, an imposed condition of fixed external source charges of the electric field in Eq.~\ref{eq:lafave-EdotD}\cite{jackson1999} is justified within a ``closed'' (connected) device as metallic electrodes rigidly define equipotential surfaces from which capacitance is derived. Moreover, electrodes consist of many free electrons, though with today's nanowires and the like, presumption of fixed external source charges is no longer merited. Absent metallic electrodes, as with biological entities for instance, for an ``open'' (isolated) device, images of internal electrons play the role of external source charges. Consequently, energy associated with image displacement must be taken into account -- the equivalent of one-half the total direct polarization energy. This constitutes fully one-half the energy stored in an isolated single electron device.

As $N\to \infty$, the direct polarization term approaches a negligible fraction of the total stored energy as justified by the approximation $dq\to 0$. The present model applies to a general $N$ electron system. Its merit is significant to the treatment of electrostatic properties of few-electron devices. 

The summation formulation provided by the discrete charge dielectric model is amenable to numerical solution. The total stored energy of a system, Eq.~\ref{eq:lafave-un}, is minimized to obtain electrons in their final electrostatic locations. Used in the treatment of spherical quantum dots, the discrete charge dielectric model has played a centrol role in the identification of a classical electrostatic ``fingerprint'' of the periodic table of elements.\cite{LaFaveJr2008-617, LaFaveJr-Dissertation} The ``fingerprint'' consists of a unique non-uniform distribution of electrostatic potential energy intrinsically associated  with point-charge symmetries of neighboring $N$-electron systems subject to spherical constraints (e.g.~a central field or a spherical quantum dot). These non-uniformities are consistent with changes in inter-atomic subshell orbital geometries, such as between the spherical $s$-orbital of an outer electron in beryllium ($Z=4$) and the energy associated with the dumbbell-shaped $p$-orbital of neighboring boron ($Z=5$) as may be empirically characterized by their respective scale-independent ionization energies. Outside of this work, classical electrostatic shell-filling behavior has not been observed in treatments of spherical quantum dots, or ``artificial atoms,'' though at least one unsuccessful attempt is described in the literature.\cite{PhysRevB.59.13036}


\newpage

\end{document}